\documentclass[english]{article}
\usepackage[T1]{fontenc}
\usepackage[latin9]{inputenc}
\usepackage{geometry}
\usepackage{float}
\usepackage{stmaryrd}
\usepackage{setspace}

\makeatletter

\providecommand{\tabularnewline}{\\}

\newcommand{\lyxaddress}[1]{
\par {\raggedright #1
\vspace{1.4em}
\noindent\par}
}

\makeatother

\usepackage{babel}
\begin{document}
\begin{doublespace}

\title{Large Changes of Atmospheric Mixing Angle $\theta_{23}$ with Four
Flavor Mixing from Planck Scale Effects}
\end{doublespace}

\author{Bipin Singh Koranga$^{1}$ and Vivek Kumar Nautiyal$^{2}$}
\maketitle

\lyxaddress{$^{1}$Department of Physics, Kirori Mal college (University of Delhi,)
Delhi-110007, India.}

\lyxaddress{$^{2}$Department of Physics, Babasaheb Bhimrao Ambedkar University,
Lucknow-226025, India}
\begin{abstract}
\begin{doublespace}
We consider the effects of Planck scale on four flavour neutrino mixings.
The gravational interaction at $M_{x}=M_{planck},$ we find that for
degenerate neurino mass order, the Planck scale effects changes the
mixing angle $\theta_{23}^{'},\,\theta_{12\,}^{'}$values and $\,\theta_{13}^{'},\,\theta_{14}^{'},\,$$\theta_{34}^{'},\,\theta_{24\,}^{'}$
are unchanged above the GUT scale. In this paper, we study,neutrino
mixing in four flavor above the GUT scale.
\end{doublespace}
\end{abstract}
\begin{doublespace}

\section{Introduction}
\end{doublespace}

Neutrino research has resulted in some of the most significant advancements
in particle physics during the last several decades. Despite this,
we still don't know everything there is to know about neutrinos. Explaining
the pattern of neutrino masses and mixings acquired from current neutrino
data is one of the most challenging challenges in neutrino physics.
The existence of non zero mass of the neutrino explained the neutrino
oscillation. Neutrino experiments like solar, atmospheric, accelerator
and reactor experiment result {[}1{]}, shows the existence of neutrino
oscillation and thus for neutrino masses and mixing. The presence
of neutrino mass and mixings is experimentally well established, subsequently
the theoretical understanding of these amounts is one of the foremost
imperative issues for particle physics. The reactor neutrino experiment
{[}2,3,4{]} discovered the unknown mixing angle $\,\theta_{13}\,$in
the three neutrino mixing frame work. The presence of sterile neutrino
points in neutrino physics. The presence of one or two sterile neutrino
states is indicated by the LSND and MiniBooNE findings. {[}5-6{]},
which adequately gives an sign to go for phenomological clarification
of four flavor system.. Neutrinos, propagating in gravitational field
has be discussed by many author {[}7{]}. In three flavor oscillation
the three mixing angles $\,\theta_{12\,},\,\theta_{23}\,$ and $\,\theta_{23}$.
In the case of four neutrino scheme (3+1), there are three extra mixing
angles $\,\theta_{14},\,\theta_{24}\,$ and $\,\theta_{34}$ . In
this paper, we consider a four flavor frame work for neutrino mixing,
where an extra sterile neutrino is introduced with the three lepton
of active neutrinos and study the all four flavor mixing angle above
the GUT scale region. The outline of the article is as follows. In
Section 2, we point out the four flavour neutrino mixing and four
flavor neutrino mixing in Planck scale region. In Section 3 and Section
4, we discuss, about numerical results and conclusions. 

\section{Four Flavour Neutrino Mixing}

In this section, we consider the (3+1) active sterile neutrino mixing,
where sterile neutrino added in mixing with three active neutrinos.
An extra sterile neutrino of small mass is introduced and it mixes
with three neutrinos. In the presence of one sterile neutrinos {[}9{]},
the PMNS matrix $U_{4\vartimes4}$ depends on the mixing angles and
CP violating phases. 

\begin{equation}
U=R_{34}R_{24}R_{14}R_{23}R_{13}R_{12}P,
\end{equation}

where the rotations matrix $R_{ij}$ is given,

\[
R_{14}=\left(\begin{array}{cccc}
c_{14} & 0 & 0 & s_{14}\\
0 & 1 & 0 & 0\\
0 & 0 & 1 & 0\\
-s_{14} & 0 & 0 & c_{14}
\end{array}\right)\,\,\,\,\,\,\,\,\,and\,\,\,\,\,\,\,\,\,R_{34}=\left(\begin{array}{cccc}
1 & 0 & 0 & 0\\
0 & 1 & 0 & 0\\
0 & 0 & c_{34} & s_{34}\\
0 & 0 & -s_{34} & c_{34}
\end{array}\right),
\]

where $s_{ij}=sin\theta_{ij}$,$\,c_{ij}=cos\theta_{ij}$. The diagonal
matrix P is function of the three Majorana phases $\alpha,\beta\,\,\,and\,\,\,\gamma:$

\begin{equation}
diag(e^{i\alpha},e^{i\beta},e^{i\gamma},1).
\end{equation}

In four flavor, there exist three Dirac phase $\delta_{ij}.$The explict
form of U is

\begin{doublespace}
\begin{equation}
\mathrm{U=\left(\begin{array}{cccc}
c_{14}c_{13}c_{12} & c_{14}c_{13}s_{12} & c_{14}s_{13} & s_{14}\\
\\
-c_{24}c_{23}s_{12} & c_{24}c_{23}c_{12} & c_{24}s_{23}c_{13} & s_{24}c_{14}\\
-c_{24}s_{23}s_{13}c_{12} & -c_{24}s_{23}s_{13}s_{12} & -s_{24}s_{14}s_{13}\\
-s_{24}s_{14}c_{13}c_{12} & -s_{24}s_{14}c_{13}s_{12}\\
\\
c_{34}s_{23}s_{12} & -c_{34}s_{23}c_{12} & c_{34}c_{23}c_{13} & s_{34}c_{24}c_{14}\\
-c_{34}c_{23}s_{13}c_{12} & -c_{34}c_{23}s_{13}s_{12} & -s_{34}s_{24}s_{23}c_{13}\\
+s_{34}s_{24}c_{13}s_{12} & -s_{34}s_{24}c_{23}c_{12} & -s_{34}c_{24}s_{14}s_{13}\\
+s_{34}s_{24}s_{23}s_{13}c_{12} & +s_{34}s_{24}s_{23}s_{13}s_{12}\\
-s_{34}c_{24}s_{14}c_{13}c_{12} & -s_{34}c_{24}s_{14}c_{13}s_{12}\\
\\
-s_{34}s_{23}s_{12} & s_{34}s_{23}c_{12} & -s_{34}c_{23}c_{13} & c_{34}c_{24}c_{14}\\
+s_{34}c_{23}s_{13}c_{12} & +s_{34}c_{23}s_{13}s_{12} & -c_{34}s_{24}s_{23}c_{13}\\
+c_{34}s_{24}c_{23}s_{12} & -c_{34}s_{24}c_{23}c_{12} & -c_{34}c_{24}s_{14}s_{13}\\
+c_{34}s_{24}s_{23}s_{12}c_{12} & +c_{34}s_{24}s_{23}s_{12}s_{12}\\
-c_{34}c_{24}s_{14}c_{13}c_{12} & -c_{34}c_{24}s_{14}c_{13}s_{12}
\end{array}\right)}\times P.
\end{equation}

\end{doublespace}

On the basis of above mixing matrix, we define given mixing angle
are

\begin{equation}
sin^{2}\mathrm{\theta_{14}=\left|U_{e4}\right|^{2}}
\end{equation}

\begin{equation}
\mathrm{sin^{2}\mathrm{\theta_{24}=\frac{\left|U_{\mu4}\right|^{2}}{1-\left|U_{e4}\right|^{2}}}}
\end{equation}

\begin{equation}
sin^{2}\mathrm{\theta_{34}=\frac{\left|U_{\tau4}\right|^{2}}{1-\left|U_{e4}\right|^{2}-\left|U_{\mu4}\right|^{2}}}
\end{equation}

\begin{equation}
\mathrm{sin^{2}\mathrm{\theta_{13}=\frac{\left|U_{e3}\right|^{2}}{1-\left|U_{e4}\right|^{2}}}}
\end{equation}

\begin{equation}
sin^{2}\mathrm{\theta_{12}=\frac{\left|U_{e2}\right|^{2}}{1-\left|U_{e4}\right|^{2}-\left|U_{e3}\right|^{2}}}
\end{equation}

\[
sin^{2}\mathrm{\theta_{23}=\frac{\left|U_{\mu3}\right|^{2}\left(1-\left|U_{e4}\right|^{2}\right)-\left(\left|U_{e4}\right|^{2}\left|U_{\mu4}\right|^{2}\right)}{1-\left|U_{e4}\right|^{2}-\left|U_{\mu4}\right|^{2}}}
\]

\begin{equation}
+\frac{\left|U_{e1}U_{\mu1}+U_{e2}U_{\mu2}\right|^{2}\left(1-\left|U_{e4}\right|^{2}\right)}{\left(1-\left|U_{e4}\right|^{2}-\left|U_{e3}\right|^{2}\right)\left(1-\left|U_{e4}\right|^{2}-\left|U_{\mu4}\right|^{2}\right)}
\end{equation}
In an earlier paper {[}10,11{]}, the neutrino mass matrix $M$~is
given by

\begin{doublespace}
\begin{equation}
\mathbf{M}=U^{*}diag(M_{i})U^{\dagger},
\end{equation}

where, $U_{\alpha i}$ and $M_{i}$, is neutrino mixing and the neutrino
masses is given by GUT. 
\end{doublespace}

The four flavor mixing for neutrinos is

\begin{doublespace}
\[
U=diag(e^{if1},e^{if2},e^{if3},e^{if4})R(\theta_{13})R(\theta_{12})R(\theta_{34})R(\theta_{24})\Delta R(\theta_{14})
\]

\end{doublespace}

\begin{equation}
\Delta^{*}R(\theta_{23})R(\theta_{34})R(\theta_{24})diag(e^{i\alpha},e^{i\beta},e^{i\gamma},1).
\end{equation}

Where

$\delta$ ,$\alpha$,~$\beta\,$and $\mathrm{\gamma}$ are the Dirac
and Majorana phases. $f_{1},$ $f_{2}$, $f_{3\,}$and $f_{4\,}$
are the phases of charge lepton field. Above the GUT scale , the effective
modified mixing matrix is {[}10{]}

\begin{doublespace}
\[
U^{'}=U(1+i\delta\theta),
\]

\end{doublespace}

\[
\mathrm{=\left(\begin{array}{cccc}
U_{e1} & U_{e2} & U_{e3} & U_{e4}\\
U_{\mu1} & U_{\mu2} & U_{\mu3} & U_{\mu4}\\
U_{\tau1} & U_{\tau2} & U_{\tau3} & U_{\tau4}\\
U_{s1} & U_{s2} & U_{s3} & U_{s4}
\end{array}\right)}
\]

\begin{equation}
+i\mathrm{\left(\begin{array}{cccc}
\mathop{\sum_{i}U_{ei}}\delta\theta_{i1} & \mathop{\sum_{i}U_{ei}}\delta\theta_{i2} & \mathop{\sum_{i}U_{ei}}\delta\theta_{i3} & \mathop{\sum_{i}U_{ei}}\delta\theta_{i4}\\
\mathop{\sum_{i}U_{\mu i}}\delta\theta_{i1} & \mathop{\sum_{i}U_{\mu i}}\delta\theta_{i2} & \mathop{\sum_{i}U_{\mu i}}\delta\theta_{i3} & \mathop{\sum_{i}U_{\mu i}}\delta\theta_{i4}\\
\mathop{\sum_{i}U_{\tau i}}\delta\theta_{i1} & \mathop{\sum_{i}U_{\tau i}}\delta\theta_{i2} & \mathop{\sum_{i}U_{\tau i}}\delta\theta_{i3} & \mathop{\sum_{i}U_{\tau i}}\delta\theta_{i4}\\
\mathop{\sum_{i}U_{si}}\delta\theta_{i1} & \mathop{\sum_{i}U_{si}}\delta\theta_{i2} & \mathop{\sum_{i}U_{si}}\delta\theta_{i3} & \mathop{\sum_{i}U_{si}}\delta\theta_{i4}
\end{array}\right)}
\end{equation}

\begin{doublespace}
Where $\delta\theta$ is a first order hermition matrix {[}10{]}.
The first order modified mass square difference {[}10{]} as

\begin{equation}
\Delta M_{ij}^{'^{2}}=\Delta M_{ij}^{2}+2(M_{i}Re(m_{ii})-M_{j}Re(m_{jj})),
\end{equation}

and
\end{doublespace}

\begin{equation}
m=\mu U^{t}\lambda U,
\end{equation}

and

\begin{doublespace}
\[
\mu=\frac{v^{2}}{M_{pl}}=2.5\times10^{-6}eV.
\]

We parametrized change in mixing angles is$\,\delta\theta\,${[}9{]}

\begin{equation}
\delta\theta_{ij}=\frac{-Im(m_{ij})(M_{i}-M_{j})+iRe(m_{ij})(M_{i}+M_{j})}{\Delta M_{ij}^{'^{2}}}.
\end{equation}

Using Eq. (12), above GUT scale the modified mixing angle is
\end{doublespace}

\begin{equation}
sin^{2}\mathrm{\theta_{14}^{'}=\left|U_{e4}^{'}\right|^{2}},
\end{equation}

\begin{equation}
\mathrm{sin^{2}\mathrm{\theta_{24}^{'}=\frac{\left|U_{\mu4}^{'}\right|^{2}}{1-\left|U_{e4}^{'}\right|^{2}},}}
\end{equation}

\begin{equation}
sin^{2}\mathrm{\theta_{34}^{'}=\frac{\left|U_{\tau4}^{'}\right|^{2}}{1-\left|U_{e4}^{'}\right|^{2}-\left|U_{\mu4}^{'}\right|^{2}}},
\end{equation}

\begin{equation}
\mathrm{sin^{2}\mathrm{\theta_{13}^{'}=\frac{\left|U_{e3}^{'}\right|^{2}}{1-\left|U_{e4}^{'}\right|^{2}}}},
\end{equation}

\begin{equation}
sin^{2}\mathrm{\theta_{12}^{'}=\frac{\left|U_{e2}^{'}\right|^{2}}{1-\left|U_{e4}^{'}\right|^{2}-\left|U_{e3}^{'}\right|^{2}}},
\end{equation}

\[
sin^{2}\mathrm{\theta_{23}^{'}=\frac{\left|U_{\mu3}^{'}\right|^{2}\left(1-\left|U_{e4}^{'}\right|^{2}\right)-\left(\left|U_{e4}^{'}\right|^{2}\left|U_{\mu4}^{'}\right|^{2}\right)}{1-\left|U_{e4}^{'}\right|^{2}-\left|U_{\mu4}^{'}\right|^{2}}}
\]

\begin{equation}
+\frac{\left|U_{e1}^{'}U_{\mu1}^{'}+U_{e2}^{'}U_{\mu2}^{'}\right|^{2}\left(1-\left|U_{e4}^{'}\right|^{2}\right)}{\left(1-\left|U_{e4}^{'}\right|^{2}-\left|U_{e3}^{'}\right|^{2}\right)\left(1-\left|U_{e4}^{'}\right|^{2}-\left|U_{\mu4}^{'}\right|^{2}\right)}.
\end{equation}

where,

\begin{equation}
\mathrm{}U_{\alpha j}^{'}=U_{\alpha j}+\mathop{\sum_{i}U_{\alpha i}}\left(\frac{-Re(m_{ij})(M_{i}+M_{j})-iIm(m_{ij})(M_{i}-M_{j})}{M_{i}^{2}-M_{j}^{2}+2(M_{i}Re(m_{ii})-M_{1}Re(m_{jj})}\right),
\end{equation}

Thus from eq.(12) and eq.(22) the deviation can be define as,
\begin{equation}
\triangle U_{\alpha j}=U_{\alpha j}^{'}-U_{\alpha j}=\mathop{i\sum_{i=1}^{4}U_{\alpha i}}\delta\theta_{ij}
\end{equation}
For denerate neutrinos mass$\,M_{i}+M_{j}\gg M_{i}-M_{j}\,$and taking
only first order approximation in $\,\mu\,$, we can get, 
\begin{equation}
\triangle U_{\alpha j}=\sum_{i}U_{\alpha i}\left(\frac{-Re(m_{ij})(M_{i}+M_{j})}{M_{i}^{2}-M_{j}^{2}}\right)\Longrightarrow\sum_{i}U_{\alpha i}\left(\frac{-Re(m_{ij})}{M_{i}-M_{j}}\right)
\end{equation}

and from eq. (14) with $\,\lambda\,$is a $\,4\times4\,$unit matrix,
we can define $\,m_{ij}\,,$ 
\[
m_{ij}=\mu m_{i}m_{j}
\]
where,
\[
m_{i}=\sum_{\alpha=e,\mu,\tau,s}U_{\alpha i}\,\,\,\,and\,\,\,\,m_{j}=\sum_{\alpha=e,\mu,\tau,s}U_{\alpha j}
\]
Thus, 

\[
\triangle U_{\alpha1}=\mu\left(U_{\alpha2}\frac{-Re(m_{21})}{M_{2}-M_{1}}+U_{\alpha3}\frac{-Re(m_{31})}{M_{3}-M_{1}}+U_{\alpha4}\frac{-Re(m_{41})}{M_{4}-M_{1}}\right)
\]
\[
\triangle U_{\alpha2}=\mu\left(U_{\alpha1}\frac{Re(m_{21})}{M_{2}-M_{1}}+U_{\alpha3}\frac{-Re(m_{32})}{M_{3}-M_{2}}+U_{\alpha4}\frac{-Re(m_{42})}{M_{4}-M_{2}}\right)
\]
\[
\triangle U_{\alpha3}=\mu\left(U_{\alpha1}\frac{Re(m_{31})}{M_{3}-M_{1}}+U_{\alpha2}\frac{Re(m_{32})}{M_{3}-M_{2}}+U_{\alpha4}\frac{Re(m_{43})}{M_{4}-M_{3}}\right)
\]
\[
\triangle U_{\alpha4}=\mu\left(U_{\alpha1}\frac{Re(m_{41})}{M_{4}-M_{1}}+U_{\alpha2}\frac{Re(m_{42})}{M_{4}-M_{2}}+U_{\alpha4}\frac{Re(m_{43})}{M_{4}-M_{3}}\right)
\]
For degenerate neutrinos masses$\,M_{4}-M_{1}\gg\gg M_{3}-M_{1}\approx M_{3}-M_{2}\gg M_{2}-M_{1\,}$because
of$\,m_{41}^{2}\gg\gg m_{31}^{2}\sim m_{32}^{2}\gg m_{21}^{2}.\,$
Since, from above equations we can conclude that the contribution
of deviation from plank scale effect for $\,\triangle U_{\alpha1}^{'},\,\triangle U_{\alpha2}^{'}\,$are
much larger than the $\,\triangle U_{\alpha3}^{'},\,\triangle U_{\alpha4}^{'}\,$
and so contribute to the values of mixing angles. Thus, we can write
the effected mixing angles$\,\theta_{23}^{'}\,\,\,\,and\,\,\,\,\theta_{12}^{'}\,$as,
\[
\theta_{23}^{'}\approx Arcsin\left(U_{e1}^{'}U_{\mu1}^{'}+U_{e2}^{'}U_{\mu2}^{'}\right)\,\,\,\,\,\,\,and\,\,\,\,\,\,\,\,\theta_{12}^{'}\approx Arcsin\left(U_{e2}^{'}\right)
\]
Therefore, we can expect a large deviation of $\,\theta_{23}\,$ and
small deviation in $\,\theta_{12}\,$ compared to other angles$\,[11]$,
while the other angles does not effected i.e $\,\theta_{14}^{'}\sim\theta_{14,}\,\theta_{24}^{'}\sim\theta_{24,}\,\theta_{34}^{'}\sim\theta_{34,}\,and\,\theta_{13}^{'}\sim\theta_{13,\,}\,.$
Above the GUT scale all six mixing angle in four flavor framework
is depend on majorona phases $\alpha,\beta$ and $\gamma$ . The modified
mixing angles are given in numerical results.
\begin{doublespace}

\section{Numerical Results}
\end{doublespace}

In numerical calculation, we consider 2eV is common neutrino mass,
which is the upper limit of tritium beta decay spectrum {[}12{]}.
We have taken $\Delta_{31}=0.002eV^{2}\,${[}13{]} , $\Delta_{21}=0.00008eV^{2}\,${[}14{]}
and $\Delta_{41}=1.7eV^{2}\,[15]$ . For simplicity, we have set the
charge lepton phases and Dirac phase $\delta=0$ and $f_{1}=f_{2}=f_{3}=f_{4}=0$.
The active sterile neutrino mixing angle are $\theta_{14},\,\,\theta_{24}\,\,and\,\,\theta_{34}$.
In this work, we consider following value for sterile neutrino mixing
angles {[}16{]}, $\theta_{14}=3.6^{o},\,\,\,\,\,\,\,\,\,\,\theta_{24}=4^{o},\,\,\,\,\,\,\,\,\,\,\theta_{34}=18.5^{o}.\,$In
Table (1.0) to Table (5.0), we take the modified mixing angle for
different sample of Majorona phases $\alpha,\,\beta$ and $\gamma$.
As shown in the table the larger deviation in $\theta_{23}\,\,\,and\,\,\,\theta_{12}$
whereas the deviation in $\theta_{14},\theta_{24},\theta_{13}$ and
$\theta_{24}$ is negligible. We see that the larger possible deviation
in $\theta_{23\,}$is about $36^{0}$ and $\theta_{12\,}$is about
$2.0^{o}.$

\begin{table}[H]
\begin{tabular}{cccccccc}
\hline 
$\beta$ & $\gamma$ & $\theta_{12}^{'}$ & $\theta_{23}^{'}$ & $\theta_{13}^{'}$ & $\theta_{14}^{'}$ & $\theta_{24}^{'}$ & $\theta_{34}^{'}$\tabularnewline
\hline 
$0^{o}$ & $0^{o}$ & 35.917 & 45.08 & 10.254 & 3.60 & 4.00 & 18.500\tabularnewline
$0^{o}$ & $45^{o}$ & 35.923 & 45.041 & 10.129 & 3.60 & 4.00 & 18.499\tabularnewline
$0^{o}$ & $90^{o}$ & 35.929 & 45.002 & 9.999 & 3.60 & 4.00 & 18.499\tabularnewline
$0^{o}$ & $135^{o}$ & 35.923 & 45.041 & 10.128 & 3.60 & 4.00 & 18.500\tabularnewline
$0^{o}$ & $180^{o}$ & 35.917 & 45.079 & 10.254 & 3.60 & 4.00 & 18.500\tabularnewline
$45^{o}$ & $0^{o}$ & 35.012 & 59.374 & 10.197 & 3.60 & 4.00 & 18.500\tabularnewline
$45^{o}$ & $45^{o}$ & 35.019 & 59.263 & 10.070 & 3.60 & 4.00 & 18.500\tabularnewline
$45^{o}$ & $90^{o}$ & 35.025 & 59.300 & 9.942 & 3.60 & 4.00 & 18.500\tabularnewline
$45^{o}$ & $135^{o}$ & 35.018 & 59.412 & 10.072 & 3.60 & 4.00 & 18.500\tabularnewline
$45^{o}$ & $180^{o}$ & 35.012 & 59.374 & 10.197 & 3.60 & 4.00 & 18.500\tabularnewline
$90^{o}$ & $0^{o}$ & 34.017 & 81.625 & 10.139 & 3.60 & 3.99 & 18.500\tabularnewline
$90^{o}$ & $45^{o}$ & 34.023 & 81.567 & 10.013 & 3.60 & 4.00 & 18.500\tabularnewline
$90^{o}$ & $90^{o}$ & 34.030 & 81.448 & 9.884 & 3.60 & 4.00 & 18.500\tabularnewline
$90^{o}$ & $135^{o}$ & 34.023 & 81.537 & 10.013 & 3.60 & 4.00 & 18.500\tabularnewline
$90^{o}$ & $180^{o}$ & 34.017 & 81.625 & 10.140 & 3.60 & 4.00 & 18.500\tabularnewline
$135^{o}$ & $0^{o}$ & 35.012 & 59.374 & 10.197 & 3.60 & 4.00 & 18.500\tabularnewline
$135^{o}$ & $45^{o}$ & 35.018 & 59.412 & 10.072 & 3.60 & 4.00 & 18.500\tabularnewline
$135^{o}$ & $90^{o}$ & 35.025 & 59.300 & 9.942 & 3.60 & 4.00 & 18.500\tabularnewline
$135^{o}$ & $135^{o}$ & 35.019 & 59.263 & 10.070 & 3.60 & 4.00 & 18.500\tabularnewline
$135^{o}$ & $180^{o}$ & 35.012 & 59.374 & 10.197 & 3.60 & 4.00 & 18.500\tabularnewline
$180^{o}$ & $0^{o}$ & 35.917 & 45.079 & 10.254 & 3.60 & 4.00 & 18.500\tabularnewline
$180^{o}$ & $45^{o}$ & 35.923 & 45.041 & 10.129 & 3.60 & 4.00 & 18.500\tabularnewline
$180^{o}$ & $90^{o}$ & 35.929 & 45.002 & 9.999 & 3.60 & 4.00 & 18.500\tabularnewline
$180^{o}$ & $135^{o}$ & 35.923 & 45.041 & 10.128 & 3.60 & 4.00 & 18.500\tabularnewline
$180^{o}$ & $180^{o}$ & 35.917 & 45.079 & 10.254 & 3.60 & 4.00 & 18.500\tabularnewline
\hline 
\end{tabular}

\caption{For $\alpha=0^{o}$, the modified mixing angles for different Majorona
phase. We take mass square difference values are $\Delta_{31}=2.0\times10^{-3}eV^{2}$,
$\Delta_{21}=8.0\times10^{-5}eV^{2}$, $\Delta_{41}=1.7eV^{2}$ and
mixing angles $\theta_{13}=10^{o},\theta_{23}=45^{o},\theta_{12}=34^{o}$
,$\theta_{34}=18.5^{o},\theta_{24}=4^{o},\theta_{14}=3.6^{o}$ }
\end{table}

\begin{table}[H]
\begin{tabular}{cccccccc}
\hline 
$\beta$ & $\gamma$ & $\theta_{12}^{'}$ & $\theta_{23}^{'}$ & $\theta_{13}^{'}$ & $\theta_{14}^{'}$ & $\theta_{24}^{'}$ & $\theta_{34}^{'}$\tabularnewline
\hline 
$0^{o}$ & $0^{o}$ & 34.922 & 59.376 & 10.184 & 3.60 & 4.00 & 18.500\tabularnewline
$0^{o}$ & $45^{o}$ & 34.928 & 59.414 & 10.058 & 3.60 & 4.00 & 18.500\tabularnewline
$0^{o}$ & $90^{o}$ & 34.935 & 59.302 & 9.929 & 3.60 & 4.00 & 18.499\tabularnewline
$0^{o}$ & $135^{o}$ & 34.929 & 59.266 & 10.059 & 3.60 & 4.00 & 18.500\tabularnewline
$0^{o}$ & $180^{o}$ & 34.922 & 59.038 & 10.184 & 3.60 & 4.00 & 18.500\tabularnewline
$45^{o}$ & $0^{o}$ & 33.993 & 45.039 & 10.128 & 3.60 & 4.00 & 18.500\tabularnewline
$45^{o}$ & $45^{o}$ & 34.000 & 45.000 & 9.999 & 3.60 & 4.00 & 18.500\tabularnewline
$45^{o}$ & $90^{o}$ & 34.007 & 44.961 & 9872 & 3.60 & 4.00 & 18.500\tabularnewline
$45^{o}$ & $135^{o}$ & 34.000 & 45.000 & 10.003 & 3.60 & 4.00 & 18.500\tabularnewline
$45^{o}$ & $180^{o}$ & 33.993 & 45.039 & 10.128 & 3.60 & 4.00 & 18.500\tabularnewline
$90^{o}$ & $0^{o}$ & 33.024 & 59.318 & 10.070 & 3.60 & 4.00 & 18.500\tabularnewline
$90^{o}$ & $45^{o}$ & 33.030 & 59.208 & 9.943 & 3.60 & 4.00 & 18.500\tabularnewline
$90^{o}$ & $90^{o}$ & 33.038 & 59.245 & 9.814 & 3.60 & 3.99 & 18.500\tabularnewline
$90^{o}$ & $135^{o}$ & 33.032 & 59.356 & 9.944 & 3.60 & 4.00 & 18.500\tabularnewline
$90^{o}$ & $180^{o}$ & 33.024 & 59.318 & 10.069 & 3.60 & 4.00 & 18.500\tabularnewline
$135^{o}$ & $0^{o}$ & 34.040 & 81.876 & 10.127 & 3.60 & 4.00 & 18.500\tabularnewline
$135^{o}$ & $45^{o}$ & 34.046 & 81.787 & 10.000 & 3.60 & 4.00 & 18.500\tabularnewline
$135^{o}$ & $90^{o}$ & 34.054 & 81.694 & 9.9871 & 3.60 & 4.00 & 18.500\tabularnewline
$135^{o}$ & $135^{o}$ & 34.048 & 81.784 & 10.001 & 3.60 & 4.00 & 18.500\tabularnewline
$135^{o}$ & $180^{o}$ & 34.040 & 81.876 & 10.127 & 3.60 & 4.00 & 18.500\tabularnewline
$180^{o}$ & $0^{o}$ & 34.922 & 59.376 & 10.184 & 3.60 & 4.00 & 18.500\tabularnewline
$180^{o}$ & $45^{o}$ & 34.928 & 59.414 & 10.058 & 3.60 & 4.00 & 18.500\tabularnewline
$180^{o}$ & $90^{o}$ & 34.935 & 59.302 & 9.929 & 3.60 & 4.00 & 18.500\tabularnewline
$180^{o}$ & $135^{o}$ & 34.929 & 59.266 & 10.059 & 3.60 & 4.00 & 18.500\tabularnewline
$180^{o}$ & $180^{o}$ & 34.922 & 59.376 & 10.184 & 3.60 & 4.00 & 18.500\tabularnewline
\hline 
\end{tabular}

\caption{For $\alpha=45^{o}$, the modified mixing angles for different Majorona
phase. We take mass square difference values are $\Delta_{31}=2.0\times10^{-3}eV^{2}$,
$\Delta_{21}=8.0\times10^{-5}eV^{2}$, $\Delta_{41}=1.7eV^{2}$ and
mixing angles $\theta_{13}=10^{o},\theta_{23}=45^{o},\theta_{12}=34^{o}$
,$\theta_{34}=18.5^{o},\theta_{24}=4^{o},\theta_{14}=3.6^{o}$ }
\end{table}

\begin{table}[H]
\begin{tabular}{cccccccc}
\hline 
$\beta$ & $\gamma$ & $\theta_{12}^{'}$ & $\theta_{23}^{'}$ & $\theta_{13}^{'}$ & $\theta_{14}^{'}$ & $\theta_{24}^{'}$ & $\theta_{34}^{'}$\tabularnewline
\hline 
$0^{o}$ & $0^{o}$ & 33.970 & 81.646 & 10.115 & 3.60 & 4.00 & 18.500\tabularnewline
$0^{o}$ & $45^{o}$ & 33.977 & 81.557 & 9.988 & 3.60 & 4.00 & 18.500\tabularnewline
$0^{o}$ & $90^{o}$ & 33.983 & 81.468 & 9.829 & 3.60 & 4.00 & 18.499\tabularnewline
$0^{o}$ & $135^{o}$ & 33.977 & 81.557 & 9.988 & 3.60 & 4.00 & 18.500\tabularnewline
$0^{o}$ & $180^{o}$ & 33.970 & 81.656 & 10.115 & 3.60 & 4.00 & 18.500\tabularnewline
$45^{o}$ & $0^{o}$ & 33.064 & 59.319 & 10.057 & 3.60 & 4.00 & 18.500\tabularnewline
$45^{o}$ & $45^{o}$ & 33.072 & 59.357 & 9.930 & 3.60 & 4.00 & 18.500\tabularnewline
$45^{o}$ & $90^{o}$ & 33.079 & 59.246 & 9.800 & 3.60 & 4.00 & 18.500\tabularnewline
$45^{o}$ & $135^{o}$ & 33.071 & 59.210 & 9.932 & 3.60 & 4.00 & 18.500\tabularnewline
$45^{o}$ & $180^{o}$ & 33.064 & 59.319 & 10.057 & 3.60 & 4.00 & 18.500\tabularnewline
$90^{o}$ & $0^{o}$ & 32.074 & 45.002 & 9.999 & 3.60 & 4.00 & 18.500\tabularnewline
$90^{o}$ & $45^{o}$ & 32.081 & 44.964 & 9.873 & 3.60 & 4.00 & 18.500\tabularnewline
$90^{o}$ & $90^{o}$ & 32.089 & 44.925 & 9.743 & 3.60 & 4.00 & 18.500\tabularnewline
$90^{o}$ & $135^{o}$ & 32.081 & 44.934 & 9.873 & 3.60 & 4.00 & 18.500\tabularnewline
$90^{o}$ & $180^{o}$ & 32.074 & 45.002 & 9.999 & 3.60 & 4.00 & 18.500\tabularnewline
$135^{o}$ & $0^{o}$ & 33.064 & 59.319 & 10.057 & 3.60 & 4.00 & 18.500\tabularnewline
$135^{o}$ & $45^{o}$ & 33.071 & 59.210 & 9.932 & 3.60 & 4.00 & 18.500\tabularnewline
$135^{o}$ & $90^{o}$ & 33.079 & 59.246 & 9.800 & 3.60 & 4.00 & 18.500\tabularnewline
$135^{o}$ & $135^{o}$ & 33.072 & 59.357 & 9.930 & 3.60 & 4.00 & 18.500\tabularnewline
$135^{o}$ & $180^{o}$ & 33.064 & 59.319 & 10.057 & 3.60 & 4.00 & 18.500\tabularnewline
$180^{o}$ & $0^{o}$ & 33.970 & 81.646 & 10.115 & 3.60 & 4.00 & 18.500\tabularnewline
$180^{o}$ & $45^{o}$ & 33.977 & 81.557 & 9.884 & 3.60 & 4.00 & 18.500\tabularnewline
$180^{o}$ & $90^{o}$ & 33.983 & 81.468 & 9.859 & 3.60 & 4.00 & 18.500\tabularnewline
$180^{o}$ & $135^{o}$ & 33.977 & 81.557 & 9.988 & 3.60 & 4.00 & 18.500\tabularnewline
$180^{o}$ & $180^{o}$ & 33.970 & 81.646 & 10.115 & 3.60 & 4.00 & 18.500\tabularnewline
\hline 
\end{tabular}

\caption{For $\alpha=90^{o}$,\textbf{ }the modified mixing angles for various
value of phases. Input values are $\Delta_{31}=2.0\times10^{-3}eV^{2}$,
$\Delta_{21}=8.0\times10^{-5}eV^{2}$, $\Delta_{41}=1.7eV^{2}$ and
mixing angles\textbf{ $\theta_{13}=10^{o},\theta_{23}=45^{o},\theta_{12}=34^{o}$
,$\theta_{34}=18.5^{o},\theta_{24}=4^{o},\theta_{14}=3.6^{o}$} }
\end{table}

\begin{table}[H]
\begin{tabular}{cccccccc}
\hline 
$\beta$ & $\gamma$ & $\theta_{12}^{'}$ & $\theta_{23}^{'}$ & $\theta_{13}^{'}$ & $\theta_{14}^{'}$ & $\theta_{24}^{'}$ & $\theta_{34}^{'}$\tabularnewline
\hline 
$0^{o}$ & $0^{o}$ & 34.922 & 59.376 & 10.184 & 3.60 & 4.00 & 18.500\tabularnewline
$0^{o}$ & $45^{o}$ & 34.929 & 59.266 & 10.059 & 3.60 & 4.00 & 18.500\tabularnewline
$0^{o}$ & $90^{o}$ & 34.935 & 59.302 & 9.929 & 3.60 & 4.00 & 18.500\tabularnewline
$0^{o}$ & $135^{o}$ & 34.928 & 59.414 & 10.058 & 3.60 & 4.00 & 18.500\tabularnewline
$0^{o}$ & $180^{o}$ & 34.922 & 59.376 & 10.184 & 3.60 & 4.00 & 18.500\tabularnewline
$45^{o}$ & $0^{o}$ & 34.040 & 81.876 & 10.127 & 3.60 & 4.00 & 18.500\tabularnewline
$45^{o}$ & $45^{o}$ & 34.048 & 81.784 & 10.001 & 3.60 & 4.00 & 18.500\tabularnewline
$45^{o}$ & $90^{o}$ & 34.054 & 81.694 & 9.871 & 3.60 & 4.00 & 18.500\tabularnewline
$45^{o}$ & $135^{o}$ & 34.040 & 81.787 & 10.000 & 3.60 & 4.00 & 18.500\tabularnewline
$45^{o}$ & $180^{o}$ & 33.024 & 81.876 & 10.127 & 3.60 & 4.00 & 18.500\tabularnewline
$90^{o}$ & $0^{o}$ & 33.032 & 59.318 & 10.069 & 3.60 & 4.00 & 18.500\tabularnewline
$90^{o}$ & $45^{o}$ & 33.038 & 59.356 & 9.944 & 3.60 & 4.00 & 18.500\tabularnewline
$90^{o}$ & $90^{o}$ & 33.024 & 59.245 & 9.814 & 3.60 & 3.99 & 18.500\tabularnewline
$90^{o}$ & $135^{o}$ & 33.993 & 59.208 & 9.942 & 3.60 & 4.00 & 18.500\tabularnewline
$90^{o}$ & $180^{o}$ & 34.000 & 59.318 & 10.069 & 3.60 & 4.00 & 18.500\tabularnewline
$135^{o}$ & $0^{o}$ & 34.000 & 45.039 & 10.127 & 3.60 & 4.00 & 18.500\tabularnewline
$135^{o}$ & $45^{o}$ & 34.000 & 45.000 & 10.003 & 3.60 & 4.00 & 18.500\tabularnewline
$135^{o}$ & $90^{o}$ & 33.993 & 44.961 & 9.9872 & 3.60 & 4.00 & 18.500\tabularnewline
$135^{o}$ & $135^{o}$ & 34.000 & 45.000 & 9.999 & 3.60 & 4.00 & 18.500\tabularnewline
$135^{o}$ & $180^{o}$ & 33.993 & 45.039 & 10.128 & 3.60 & 4.00 & 18.500\tabularnewline
$180^{o}$ & $0^{o}$ & 34.922 & 59.376 & 10.184 & 3.60 & 4.00 & 18.500\tabularnewline
$180^{o}$ & $45^{o}$ & 34.929 & 59.266 & 10.059 & 3.60 & 4.00 & 18.500\tabularnewline
$180^{o}$ & $90^{o}$ & 34.935 & 59.302 & 9.929 & 3.60 & 4.00 & 18.500\tabularnewline
$180^{o}$ & $135^{o}$ & 34.928 & 59.414 & 10.057 & 3.60 & 4.00 & 18.500\tabularnewline
$180^{o}$ & $180^{o}$ & 34.922 & 59.376 & 10.184 & 3.60 & 4.00 & 18.500\tabularnewline
\hline 
\end{tabular}

\caption{For $\alpha=135^{o}$, the modified mixing angles for different Majorona
phase. We take mass square difference values are $\Delta_{31}=2.0\times10^{-3}eV^{2}$,
$\Delta_{21}=8.0\times10^{-5}eV^{2}$, $\Delta_{41}=1.7eV^{2}$ and
mixing angles $\theta_{13}=10^{o},\theta_{23}=45^{o},\theta_{12}=34^{o}$
,$\theta_{34}=18.5^{o},\theta_{24}=4^{o},\theta_{14}=3.6^{o}$ }
\end{table}

\begin{table}[H]
\begin{tabular}{cccccccc}
\hline 
$\beta$ & $\gamma$ & $\theta_{12}^{'}$ & $\theta_{23}^{'}$ & $\theta_{13}^{'}$ & $\theta_{14}^{'}$ & $\theta_{24}^{'}$ & $\theta_{34}^{'}$\tabularnewline
\hline 
$0^{o}$ & $0^{o}$ & 35.917 & 45.079 & 10.254 & 3.60 & 4.00 & 18.500\tabularnewline
$0^{o}$ & $45^{o}$ & 35.923 & 45.041 & 10.129 & 3.60 & 4.00 & 18.500\tabularnewline
$0^{o}$ & $90^{o}$ & 35.929 & 45.002 & 9.999 & 3.60 & 4.00 & 18.500\tabularnewline
$0^{o}$ & $135^{o}$ & 35.923 & 45.041 & 10.129 & 3.60 & 4.00 & 18.500\tabularnewline
$0^{o}$ & $180^{o}$ & 35.917 & 45.079 & 10.254 & 3.60 & 4.00 & 18.500\tabularnewline
$45^{o}$ & $0^{o}$ & 35.012 & 59.374 & 10.197 & 3.60 & 4.00 & 18.500\tabularnewline
$45^{o}$ & $45^{o}$ & 35.019 & 59.263 & 10.070 & 3.60 & 4.00 & 18.500\tabularnewline
$45^{o}$ & $90^{o}$ & 35.025 & 59.300 & 9.942 & 3.60 & 4.00 & 18.500\tabularnewline
$45^{o}$ & $135^{o}$ & 35.018 & 59.412 & 10.072 & 3.60 & 4.00 & 18.500\tabularnewline
$45^{o}$ & $180^{o}$ & 35.012 & 59.374 & 10.197 & 3.60 & 4.00 & 18.500\tabularnewline
$90^{o}$ & $0^{o}$ & 34.017 & 81.625 & 10.139 & 3.60 & 4.00 & 18.500\tabularnewline
$90^{o}$ & $45^{o}$ & 34.023 & 81.537 & 10.013 & 3.60 & 4.00 & 18.500\tabularnewline
$90^{o}$ & $90^{o}$ & 34.030 & 81.448 & 9884 & 3.60 & 3.99 & 18.500\tabularnewline
$90^{o}$ & $135^{o}$ & 34.023 & 81.537 & 10.013 & 3.60 & 4.00 & 18.500\tabularnewline
$90^{o}$ & $180^{o}$ & 34.017 & 81.625 & 10.139 & 3.60 & 4.00 & 18.500\tabularnewline
$135^{o}$ & $0^{o}$ & 35.012 & 59.374 & 10.197 & 3.60 & 4.00 & 18.500\tabularnewline
$135^{o}$ & $45^{o}$ & 35.018 & 59.412 & 10.072 & 3.60 & 4.00 & 18.500\tabularnewline
$135^{o}$ & $90^{o}$ & 35.025 & 59.300 & 9.942 & 3.60 & 4.00 & 18.500\tabularnewline
$135^{o}$ & $135^{o}$ & 35.019 & 59.263 & 10.070 & 3.60 & 4.00 & 18.500\tabularnewline
$135^{o}$ & $180^{o}$ & 35.012 & 59.374 & 10.197 & 3.60 & 4.00 & 18.500\tabularnewline
$180^{o}$ & $0^{o}$ & 35.917 & 45.079 & 10.254 & 3.60 & 4.00 & 18.500\tabularnewline
$180^{o}$ & $45^{o}$ & 35.923 & 45.041 & 10.129 & 3.60 & 4.00 & 18.500\tabularnewline
$180^{o}$ & $90^{o}$ & 35.929 & 45.002 & 9.999 & 3.60 & 4.00 & 18.500\tabularnewline
$180^{o}$ & $135^{o}$ & 35.923 & 45.041 & 10.129 & 3.60 & 4.00 & 18.500\tabularnewline
$180^{o}$ & $180^{o}$ & 35.917 & 45.079 & 10.254 & 3.60 & 4.00 & 18.500\tabularnewline
\hline 
\end{tabular}

\caption{For $\alpha=180^{o}$, the modified mixing angles for different Majorona
phase. We take mass square difference values are $\Delta_{31}=2.0\times10^{-3}eV^{2}$,
$\Delta_{21}=8.0\times10^{-5}eV^{2}$, $\Delta_{41}=1.7eV^{2}$ and
mixing angles $\theta_{13}=10^{o},\theta_{23}=45^{o},\theta_{12}=34^{o}$
,$\theta_{34}=18.5^{o},\theta_{24}=4^{o},\theta_{14}=3.6^{o}$}
\end{table}

\begin{doublespace}

\section{Conclusions}
\end{doublespace}

\begin{doublespace}
In this paper, we tested how Planck scale effects on neutrino mixing
angle in a four flavor framework. We calculate the neutrino mixing
angle above the GUT scale. On the basis of current neutrino oscillation
parameters, we have predicted the neutrino mixing angle predection
is $\theta_{12}^{Planck}=\theta_{12}+1.92^{o}$ \textbf{,}$\theta_{23}^{Planck}=\theta_{23}+36.62^{o}$
\textbf{,}$\theta_{13}^{Planck}=\theta_{13}+0.25^{o}$ \textbf{,}$\theta_{24}^{Planck}=\theta_{24}+0{}^{o}$
\textbf{,}$\theta_{34}^{Planck}=\theta_{34}+0^{o}$ and\textbf{ }$\theta_{14}^{Planck}=\theta_{14}+0{}^{o}$
, which is corresponds to Planck scale $M_{Planck}\simeq2\times10^{19}GeV.$
We have calculated the neutrino mixing angle in four flavor framework
above the GUT scale.We consider all possible range of Majorana phases,we
found above the GUT scale deviation of atmospheric mixing angle is
larger $\theta_{23}^{'}$ . There is no any changes of mixing angle
$\theta_{24}^{Planck},$ $\theta_{34}^{Planck}$ and\textbf{ }$\theta_{14}^{Planck}.$This
calculation for only degenerate neutrino mass about 2eV.
\end{doublespace}

\end{document}